\documentclass[aps,preprint,12pt,a4paper]{revtex4}%
\usepackage{amsmath}
\usepackage{graphicx}
\usepackage{amsfonts}
\usepackage{amssymb}%
\begin{document}
\preprint{quant-ph/0303119}
\title{{\Large Squeezing arbitrary cavity-field states through their interaction with
a single driven atom}}
\author{C. J. Villas-B\^{o}as$^{1}$, N. G. de Almeida$^{2}$, R. M. Serra$^{1} $, and
M. H. Y. Moussa$^{1}$}
\affiliation{$^{1}$ Departamento de F\'{\i}sica, Universidade Federal de S\~{a}o Carlos,
P.O. Box 676, S\~{a}o Carlos, 13565-905, S\~{a}o Paulo, Brazil.\linebreak%
$^{2}$ Departamento de Matem\'{a}tica e F\'{\i}sica, Universidade Cat\'{o}lica
de Goi\'{a}s, P.O. Box 86, Goi\^{a}nia,74605-010, Goi\'{a}s, Brazil.}

\begin{abstract}
We propose an implementation of the parametric amplification of an arbitrary
radiation-field state previously prepared in a high-$Q$ cavity. This nonlinear
process is accomplished through the dispersive interactions of a single
three-level atom (fundamental $\left\vert g\right\rangle $, intermediate
$\left\vert i\right\rangle $, and excited $\left\vert e\right\rangle $ levels)
simultaneously with $i)$ a classical driving field and $ii)$ a previously
prepared cavity mode whose state we wish to squeeze. We show that, in the
adiabatic approximantion, the preparation of the initial atomic state in the
intermediate level $\left\vert i\right\rangle $ becomes crucial for obtaing
the degenerated parametric amplification process.


\textbf{Journal-ref:: } Phys. Rev. A \textbf{68}, 061801(R) (2003)

\end{abstract}
\maketitle

The parametric amplification process represents a central issue in quantum
optics since its applications range from fundamental physics to technology. As
one of its by-products, the squeezed states of the radiation field have been
researched in order to deepen our understanding of the properties of radiation
\cite{Stoler} and its interaction with matter \cite{Milburn}. An unequivocal
signature of the quantum nature of light has been provided\ by the
antibunching process emerging from squeezed light, and apart from fundamental
questions, squeezed states have provoked some striking technological
challenges. The improvement of the signal to noise ratio in optical
communication \cite{SN} and, even more\ attractive, the possibility of
measuring gravitational waves through squeezed fields \cite{Caves}, are just
some of the potential applications of squeezed states. Such proposals rely on
reducing the quantum fluctuation in one (signal) quadrature component of the
field at the expense of amplifying the fluctuation in another (unobservable)
component, as ruled by the Heisenberg uncertainty relation \cite{Dodonov}.

As squeezed light is mainly supplied by nonlinear optical media as running
waves (through backward \cite{98} or forward \cite{99} four-wave mixing and
parametric down-conversion \cite{100}), standing squeezed fields in high-$Q$
cavities or ion traps can be generated through atom-field interaction
\cite{Meystre}. Although considerable space has been devoted in the literature
to the squeezing process in the Jaynes-Cummings model, the issue of squeezing
any desired prepared cavity-field state $\left|  \Psi\right\rangle $, i.e.,
the accomplishment of the operation $S(\zeta)\left|  \Psi\right\rangle $ in
cavity QED ($\zeta$ standing for a set of group parameters) has not been
addressed. Engineering such an operation is the subject of the present letter;
it is achieved through the dispersive interactions of a three-level atom
simultaneously with a classical driving field and a cavity mode whose prepared
state we wish to squeeze. In short, the dispersive interaction of the cavity
mode with a driven atom produces the desired operation $S(\zeta)\left|
\Psi\right\rangle $.

Selective atomic measurements in cavity QED have been employed to enhance
squeezing in the Jaynes-Cummings model (JCM) \cite{Gerry}. Whereas
cavity-field squeezing in the JCM is rather modest, about 20\% for low average
photon number, squeezing of up to 75\% can be obtained through selective
atomic measurements \cite{Gerry}.{\large \ }However, the squeezed states
resulting from selective atomic measurements (and the proposed schemes
employing atom-field interactions \cite{BW}) do not come from the unitary
evolution $S(\zeta)\left|  \Psi\right\rangle $. In the present proposal of
dispersive interaction of the cavity mode -- whose state $\left|
\Psi\right\rangle $ is to be squeezed -- with a driven atom, we obtain
squeezing around 88\%. This higher squeezing is crucial to the building of
truly mesoscopic superpositions with a large average photon number and also a
large ``distance'' in phase space between the centers of the quasi-probability
distribution of the individual states composing the prepared superposition
\cite{Celso}.

As depicted in Fig. 1, the three-level atom is in a ladder configuration where
an intermediate atomic level ($\left|  i\right\rangle $) lies between the
ground ($\left|  g\right\rangle $) and the excited ($\left|  e\right\rangle $)
states. The quantized cavity mode of frequency $\omega$ couples dispersively
both transitions $\left|  g\right\rangle $ $\leftrightarrow$ $\left|
i\right\rangle $ and $\left|  e\right\rangle $ $\leftrightarrow$ $\left|
i\right\rangle $ with coupling constants $\lambda_{g}$ and $\lambda_{e}$,
respectively, and detuning $\delta=\left|  \omega-\omega_{\ell i}\right|  $
($\ell=g,e$). A classical field of frequency $\omega_{0}=2\omega+\Delta$
drives dispersively the atomic transition $\left|  g\right\rangle $
$\leftrightarrow$ $\left|  e\right\rangle $ with coupling constant $\Omega
$.{\Large \ }(We assume that the transition $\left|  g\right\rangle $
$\leftrightarrow$ $\left|  e\right\rangle $ may be induced by applying a
sufficiently strong electric field) While the quantum field promotes a
two-photon interchange process, the classical driving field constitutes the
source of the parametric amplification. This system has been considered in
many theoretical \cite{Luis} and experimental works \cite{Raimond}

The Hamiltonian of our model, within the rotating wave approximation, is given
by $H=H_{0}+V$, where
\begin{subequations}
\begin{align}
H_{0}  &  =\hbar\omega a^{\dagger}a-\hbar\omega\left|  g\right\rangle
\left\langle g\right|  +\hbar\delta\left|  i\right\rangle \left\langle
i\right|  +\hbar\omega\left|  e\right\rangle \left\langle e\right|
,\label{Eq1a}\\
V  &  =\hbar\left(  \lambda_{g}a\left|  i\right\rangle \left\langle g\right|
+H.c.\right)  +\hbar\left(  \lambda_{e}a\left|  e\right\rangle \left\langle
i\right|  +H.c\right) \nonumber\\
&  +\hbar\left(  \Omega\left|  e\right\rangle \left\langle g\right|
e^{-i\omega_{0}t}+H.c.\right)  , \label{Eq1b}%
\end{align}
with $a^{\dagger}$ ($a$) standing for the creation (annihilation) operator of
the quantized cavity mode. Writing $H$ in the interaction picture (through the
unitary transformation $U_{0}=\exp\left(  -iH_{0}t/\hbar\right)  $) and then
applying the transformation\ $U=\exp\left[  -i\delta t\left(  \left|
g\right\rangle \left\langle g\right|  +\left|  e\right\rangle \left\langle
e\right|  \right)  \right]  $, we obtain the Hamiltonian $\mathbf{H}%
=U_{0}^{\dagger}U^{\dagger}HUU_{0}-H_{0}-\hbar\delta\left(  \left|
g\right\rangle \left\langle g\right|  +\left|  e\right\rangle \left\langle
e\right|  \right)  $\ given by
\end{subequations}
\begin{align}
\mathbf{H}  &  =\hbar\left(  \lambda_{g}a\sigma_{ig}+\lambda_{e}a\sigma
_{ei}+\Omega\operatorname*{e}\nolimits^{-i\Delta t}\sigma_{eg}+H.c\right)
\nonumber\\
&  -\hbar\delta\left(  \sigma_{gg}+\sigma_{ee}\right)  , \label{Eq2}%
\end{align}
where we have defined the atomic transition operator $\sigma_{kl}\equiv\left|
k\right\rangle \left\langle l\right|  $, $k,l=g,i,e$. Next, we compare the
time scales of the transitions induced by the cavity field, considering the
Heisenberg equations of motion for the transition operators $\sigma_{ig}$ and
$\sigma_{ei}$,
\begin{subequations}
\begin{align}
i\frac{d}{dt}\sigma_{ig}  &  =\lambda_{g}^{\ast}a^{\dagger}\left(  \sigma
_{ii}-\sigma_{gg}\right)  -\lambda_{e}a\sigma_{eg}+\Omega^{\ast}%
\operatorname*{e}\nolimits^{i\Delta t}\sigma_{ie}-\delta\sigma_{ig}%
\mathrm{{,}}\label{Eq3a}\\
i\frac{d}{dt}\sigma_{ei}  &  =\lambda_{g}a\sigma_{eg}+\lambda_{e}^{\ast
}a^{\dagger}\left(  \sigma_{ee}-\sigma_{ii}\right)  -\Omega^{\ast
}\operatorname*{e}\nolimits^{i\Delta t}\sigma_{gi}+\delta\sigma_{ei}%
\mathrm{{.}} \label{Eq3b}%
\end{align}
If the dispersive transitions are sufficiently{\large \ }detuned, i.e.,
$\delta\gg$ $\left|  \lambda_{g}\right|  $,$\left|  \lambda_{e}\right|
$,$\left|  \Omega\right|  $,$\left|  \Delta\right|  $, we obtain the adiabatic
solutions for the transition operators $\sigma_{ig}$ and $\sigma_{ei}$, by
setting $d\sigma_{ig}/dt=d\sigma_{ei}/dt=0$ \cite{Allen}:
\end{subequations}
\begin{subequations}
\begin{align}
\sigma_{ig}  &  =\frac{1}{\delta}\left[  \lambda_{g}^{\ast}a^{\dagger}\left(
\sigma_{ii}-\sigma_{gg}\right)  -\lambda_{e}a\sigma_{eg}+\Omega^{\ast
}\operatorname*{e}\nolimits^{i\Delta t}\sigma_{ie}\right]  ,\label{Eq4a}\\
\sigma_{ei}  &  =\frac{1}{\delta}\left[  \lambda_{e}^{\ast}a^{\dagger}\left(
\sigma_{ii}-\sigma_{ee}\right)  -\lambda_{g}a\sigma_{eg}+\Omega^{\ast
}\operatorname*{e}\nolimits^{i\Delta t}\sigma_{gi}\right]  . \label{Eq4b}%
\end{align}
Solving the system (\ref{Eq4a},\ref{Eq4b}) and inserting these adiabatic
solutions for $\sigma_{ig}$ and $\sigma_{ei}$ into Eq. (\ref{Eq2}), the
Hamiltonian becomes
\end{subequations}
\begin{align}
\mathbf{H}  &  =-\hbar\delta\left(  \sigma_{gg}+\sigma_{ee}\right)
+\hbar\left(  \Omega\operatorname*{e}\nolimits^{-i\Delta t}\sigma
_{eg}+H.c\right)  -\frac{\hbar}{\delta}\left\{  \left(  2a^{\dagger
}a+1\right)  \right. \nonumber\\
&  \times\left[  \left|  \lambda_{g}\right|  ^{2}\sigma_{gg}-\left(  \left|
\lambda_{g}\right|  ^{2}+\left|  \lambda_{e}\right|  ^{2}\right)  \sigma
_{ii}+\left|  \lambda_{e}\right|  ^{2}\sigma_{ee}+\frac{\left|  \lambda
_{g}\right|  ^{2}+\left|  \lambda_{e}\right|  ^{2}}{2\delta}\left(
\Omega\operatorname*{e}\nolimits^{-i\Delta t}\sigma_{eg}+H.c.\right)  \right]
\nonumber\\
&  \left.  +2\left(  \lambda_{g}\lambda_{e}a^{2}\sigma_{eg}+H.c.\right)
+\frac{1}{\delta}\left(  \lambda_{g}\lambda_{e}\Omega^{\ast}\operatorname*{e}%
\nolimits^{i\Delta t}a^{2}+H.c.\right)  \left(  \sigma_{gg}+\sigma
_{ee}-2\sigma_{ii}\right)  \right\}  \label{Eq5}%
\end{align}
We note that the solution of the system (\ref{Eq4a},\ref{Eq4b}) must be
inserted into a symmetrized Hamiltonian (\ref{Eq2}), where the operator
structure $a\sigma_{kl}$ must be substituted by $\left(  a\sigma_{kl}%
+\sigma_{kl}a\right)  /2$. Otherwise, the resulting Hamiltonian (\ref{Eq5})
would depend on the order of the operators, $a\sigma_{kl}$ or $\sigma_{kl}a$.
The state vector associated with Hamiltonian (\ref{Eq5}), in the
Schr\"{o}dinger picture, can be written using
\begin{equation}
|\Psi\left(  t\right)  \rangle=\left|  g\right\rangle \left|  \Phi_{g}\left(
t\right)  \right\rangle +\left|  i\right\rangle \left|  \Phi_{i}\left(
t\right)  \right\rangle +\left|  e\right\rangle \left|  \Phi_{e}\left(
t\right)  \right\rangle \mathrm{{,}} \label{Eq6}%
\end{equation}
where $|\Phi_{\ell}\left(  t\right)  \rangle=\int\frac{d^{2}\alpha}{\pi
}\mathcal{A}_{\ell}\left(  \alpha,t\right)  |\alpha\rangle$, $\ell=g,i,e$, the
complex quantity $\alpha$ standing for the eigenvalues of $a$, and
$\mathcal{A}_{\ell}\left(  \alpha,t\right)  =\left\langle \alpha,\ell\left|
\Psi\left(  t\right)  \right.  \right\rangle $ is the expansion coefficients
for $|\Phi_{\ell}\left(  t\right)  \rangle$ in the basis of coherent states,
$\left\{  |\alpha\rangle\right\}  $. Using the orthogonality of the atomic
states and Eqs. (\ref{Eq5}) and (\ref{Eq6}) we obtain the uncoupled
time-dependent (TD) Schr\"{o}dinger equations for the atomic subspace $\left|
i\right\rangle $ (in the Schr\"{o}dinger picture):
\begin{equation}
i\hbar\frac{d}{dt}|\Phi_{i}\left(  t\right)  \rangle=\mathcal{H}_{i}|\Phi
_{i}\left(  t\right)  \rangle\mathrm{{,}} \label{Eq7}%
\end{equation}%
\begin{equation}
\mathcal{H}_{i}=\hbar\varpi a^{\dagger}a+\hbar\left(  \xi\operatorname*{e}%
\nolimits^{-i\nu t}a^{\dagger^{2}}+\xi^{\ast}\operatorname*{e}\nolimits^{i\nu
t}a^{2}\right)  \label{Eq8}%
\end{equation}
where $\varpi=\omega+\chi$ $\left(  \chi=\left.  2\left(  \left|  \lambda
_{g}\right|  ^{2}+\left|  \lambda_{e}\right|  ^{2}\right)  \right/
\delta\right)  $ stands for the effective frequency of the cavity mode, while
$\xi=2\Omega\lambda_{g}^{\ast}\lambda_{e}^{\ast}/\delta^{2}=\left|
\xi\right|  \operatorname*{e}\nolimits^{-i\Theta}$ and $\nu=2\omega+\Delta$
are the effective amplitude and frequency of the parametric amplification
field. For subspace $\left\{  \left|  g\right\rangle ,\left|  e\right\rangle
\right\}  $ there is a TD Schr\"{o}dinger equation which couples the
fundamental and the excited atomic states. Therefore, when we initially
prepare the atom in the intermediate level $\left|  i\right\rangle $, the
dynamics of the atom-field dispersive interactions, governed by the effective
Hamiltonian (\ref{Eq8}), results in a cavity mode with shifted frequency
submitted to a parametric amplification process.

In the resonant regime the classical driving field has the same frequency as
the cavity mode, so that $\nu=2\varpi$ (i.e. $\Delta=2\chi$). The evolution of
the cavity field state, in the interaction picture, is governed by a squeeze
operator such as $|\Phi_{i}\left(  t\right)  \rangle=S(\xi,t)|\Phi_{i}\left(
0\right)  $, where
\begin{equation}
S(\xi,t)=\exp\left[  -i\left(  \xi a^{\dagger2}+\xi^{\ast}a^{2}\right)
t\right]  . \label{Eq9}%
\end{equation}
The degree of squeezing in the on-resonant regime is determined by the factor
$r_{on}(t)=2\left|  \xi\right|  t$, while the squeeze angle is given by
$\varphi_{on}=\pi/2-\Theta$. For a specific cavity mode and atomic
configuration, the parameter $r_{on}(t)$ can be adjusted in accordance with
the coupling strength $\Omega$ and the interaction time $t$. Assuming typical
values for the parameters involved, arising from Rydberg states where the
intermediate state $\left|  i\right\rangle $ is nearly halfway between
$\left|  g\right\rangle $ and $\left|  e\right\rangle $, we get $\left|
\lambda_{g}\right|  \sim\left|  \lambda_{e}\right|  \sim3\times10^{5}$s$^{-1}%
$\cite{Walther,Haroche01}. With such values and assuming the detuning $\left|
\delta\right|  \sim15\times\left|  \lambda_{g}\right|  $ and the coupling
strength also $\Omega\sim3\times10^{5}$s$^{-1}$, we obtain $\left|
\xi\right|  \sim3\times10^{3}$s$^{-1}$. For an atom-field interaction time
about $t\sim2\times10^{-4}$s, we get the squeezing factor $r_{on}(t)\sim1.07$
such that, for the resonant regime, the variance in the squeezed quadrature
turns out to be $\left\langle \Delta X\right\rangle ^{2}=\operatorname{e}%
^{-2r_{on}(t)}/4\sim3\times10^{-2}$, representing a squeezing around $88\%$
(for an initial coherent state prepared in the cavity) with the passage of
just one atom. Of course, the injection of more atoms through the cavity leads
to a squeezing even greater than this remarkable rate. Note that the
interaction time considered here is two (one) order of magnitude smaller than
the field decay time in closed \cite{Walther} (open \cite{Haroche01})
microwave cavities used in these experiments. We note that closed cavities do
not allow the use of circular Rydberg atoms and, consequently, the atomic
decay time becomes a concern \cite{Ruynet}.

Even that the adjustment of the detuning $\Delta$ between the driving field
and the atomic transition (such that $\Delta=2\chi$) is not the main
difficulty of implementing the method here proposed we also analyzed the
off-resonant regime ($\nu\neq2\varpi$). To solve the Schr\"{o}dinger Eq.
(\ref{Eq7}) we employed the TD invariants of Lewis and Riesenfeld \cite{Lewis}
as demonstrated in detail in \cite{Celso}. It is possible to show
\cite{Celso,Salomon} that in the off-resonant regime we find three different
solutions depending on parameter $\mathfrak{P}=4\left|  \xi\right|
/(2\chi-\Delta)$ which is an effective macroscopic coupling: the strong
($\left|  \mathfrak{P}\right|  >1$), weak ($\left|  \mathfrak{P}\right|  <1$),
and critical ($\left|  \mathfrak{P}\right|  =1$) coupling parameter. There is
a well-known threshold in the behavior of the TD squeeze factor $r_{off}%
(t)$,\ arising from the quadratic TD Hamiltonian (\ref{Eq8})
\cite{Celso,Salomon}: $r_{off}(t)$\ increases monotonically for $\left|
\mathfrak{P}\right|  \geq1$,\ while for $\left|  \mathfrak{P}\right|  <1$\ it
oscillates periodically. For this reason, in the present letter we are
interested in the strong coupling regime, where we obtain the highest TD
squeeze parameters, given by
\begin{subequations}
\begin{align}
\cosh\left(  2r_{off}(t)\right)   &  =\frac{1}{\mathfrak{P}^{2}-1}\left[
\frac{\operatorname{e}^{h(t)}}{4}+\mathfrak{P}^{2}\left(  \mathfrak{C}%
^{2}+\mathfrak{P}^{2}-1\right)  \operatorname{e}^{-h(t)}-\mathfrak{C}\right]
{.}\label{Eq10a}\\
\cos\left[  \varphi_{off}(t)+\nu t-\Theta\right]   &  =\frac{\mathfrak{C}%
-\cosh\left(  2r_{off}(t)\right)  }{\mathfrak{P}\sinh\left(  2r_{off}%
(t)\right)  }\mathrm{{,}} \label{Eq10b}%
\end{align}
where the constant $\mathfrak{C}$ and function $h(t)$ are given, respectively, by%

\end{subequations}
\begin{subequations}
\begin{align}
\mathfrak{C}  &  =\cosh\left[  2r_{off}(0)\right]  +\mathfrak{P}\cos\left[
\varphi_{off}(0)-\Theta\right]  \sinh\left[  2r_{off}(0)\right]
\label{Eq11a}\\
h(t)  &  =\mp\frac{\sqrt{\mathfrak{P}^{2}-1}}{\left|  \mathfrak{P}\right|
}4\xi t+\ln\left[  2\left|  \mathfrak{P}\right|  \left(  \sqrt{\left(
\mathfrak{P}^{2}-1\right)  \left(  \mathfrak{C}^{2}-1\right)  }+\mathfrak{C}%
\left|  \mathfrak{P}\right|  \right)  \right]  \mathrm{{,}} \label{Eq11b}%
\end{align}
the sign being chosen so that $r_{off}(t)\geq0$. We note that in the limit as
$\left|  \mathfrak{P}\right|  \rightarrow\infty$, i.e, $\Delta\rightarrow
2\chi$, we obtain the on-resonant interaction from the dispersive strong
coupling regime. Therefore, the highest squeezing factor, resulting from the
highest intensity of the effective coupling $\mathfrak{P}$, is computed from
the on-resonant regime, as observed in Fig. 2, where the ratio $r_{off}%
(t)/r_{on}(t)$ is depicted as a function of the detuning $\Delta$.

It is worth stressing that for weak damped systems, as fields trapped into
realistic high-$Q$ cavities, the lifetime of the squeezing is of order of the
relaxation time of the cavity \cite{Grabert}. Therefore, the dissipative
mechanism of the cavity plays a much milder role in the lifetime of the
squeezing than in decoherence phenomena \cite{LD}. Regarding atomic decay, we
note that for circular Rydberg levels the spontaneous emission hardly affects
the squeezing process for typical interaction time scales. In this connection,
next we estimate the on-resonant squeezing factor considering the finite
lifetime of the atomic levels as well as the cavity damping rate which are
introduced phenomenologically into the equation of motion
\end{subequations}
\begin{equation}
\frac{d}{dt}\mathcal{O}=-\frac{i}{\hbar}\left[  \mathbf{H,}\mathcal{O}\right]
-\frac{\Gamma}{2}\mathcal{O}\text{,} \label{E1}%
\end{equation}
where $\Gamma$ stands for the decay rate of the system corresponding to
operator $\mathcal{O}$ and Hamiltonian $\mathbf{H}$ is given by Eq.
(\ref{Eq2}) \cite{Scully}. To estimate the squeezing factor we compute the
variance of the field quadrature $X=\left(  a\operatorname*{e}%
\nolimits^{-i\varphi_{on}}+a^{\dagger}\operatorname*{e}\nolimits^{i\varphi
_{on}}\right)  /2$ from the solution of equation%
\begin{equation}
\frac{d}{dt}a=i\left(  \lambda_{g}\sigma_{ig}+\lambda_{e}\sigma_{ei}\right)
-\frac{\Gamma_{c}}{2}a\text{,} \label{E2}%
\end{equation}
where $\Gamma_{c}$ indicates the cavity damping rate. Proceeding to the
adiabatic solutions of equations
\begin{subequations}
\begin{align}
i\frac{d}{dt}\sigma_{ig}  &  =\lambda_{g}^{\ast}a^{\dagger}\left(  \sigma
_{ii}-\sigma_{gg}\right)  -\lambda_{e}a\sigma_{eg}+\Omega^{\ast}%
\operatorname*{e}\nolimits^{i\Delta t}\sigma_{ie}-\delta\sigma_{ig}%
-\frac{\Gamma_{i}}{2}\sigma_{ig}\text{{,}}\label{E3a}\\
i\frac{d}{dt}\sigma_{ei}  &  =\lambda_{g}a\sigma_{eg}+\lambda_{e}^{\ast
}a^{\dagger}\left(  \sigma_{ee}-\sigma_{ii}\right)  -\Omega^{\ast
}\operatorname*{e}\nolimits^{i\Delta t}\sigma_{gi}+\delta\sigma_{ei}%
-\frac{\Gamma_{e}}{2}\sigma_{ei}\text{{,}} \label{E3b}%
\end{align}
assuming now that the dispersive transitions are sufficiently{\large \ }%
detuned such that $\delta\gg$ $\left|  \lambda_{g}\right|  $,$\left|
\lambda_{e}\right|  $,$\left|  \Omega\right|  $,$\left|  \Delta\right|
$,$\Gamma_{ig}$,$\Gamma_{ei}$, we obtain the solutions in Eqs. (\ref{Eq4a})
and (\ref{Eq4b}) except for changing $\delta$ by $\delta-i\Gamma_{i}$ and
$\delta-i\Gamma_{e}$, respectively.

In what follows we consider three approximations in order to simplify our
calculations. First, $i)$ we assume the same lifetime for both atomic levels
$\left|  e\right\rangle $ and $\left|  i\right\rangle $ to define the atomic
decay rate $\Gamma_{a}=\Gamma_{i}=\Gamma_{e}$. Secondly, $ii)$ we assume that
the atomic decay will hardly populate level $\left|  g\right\rangle $ and,
consequently, level $\left|  e\right\rangle $ (which is coupled to $\left|
g\right\rangle $ through the classical field). (In fact, even in the ideal
situation where dissipation is dismissed, the experiment must be restarted
when the atom is not detected in the state $\left|  i\right\rangle $ after
interacting with the cavity field.) With this assumption, which considerably
simplify the problem, we obtain the commutation $\left[  \sigma_{ii}%
\mathbf{,H}\right]  \varpropto\sigma_{gg}$, $\sigma_{ee}$, $\sigma_{ge}$
$\approx0$ such that $\sigma_{ii}(t)=\operatorname*{e}\nolimits^{-\Gamma_{a}%
t}\sigma_{ii}(0)$. Substituting the solutions for $\sigma_{ei}$ and
$\sigma_{ig}$ (resulting from these two approximations besides the adiabatic
one) into Eq. (\ref{E2}) we finally obtain the coupled equations%
\end{subequations}
\begin{subequations}
\begin{align}
\frac{d}{dt}\widetilde{a}  &  =-i\chi(1-\operatorname*{e}\nolimits^{-\Gamma
_{a}t/2})\widetilde{a}+i\xi\operatorname*{e}\nolimits^{-\Gamma_{a}%
t/2}\widetilde{a}^{\dagger}\text{,}\label{E4a}\\
\frac{d}{dt}\widetilde{a}^{\dagger}  &  =i\chi(1-\operatorname*{e}%
\nolimits^{-\Gamma_{a}t/2})\widetilde{a}\dagger-i\xi\operatorname*{e}%
\nolimits^{-\Gamma_{a}t/2}\widetilde{a}\text{,} \label{E4b}%
\end{align}
where $\widetilde{a}=\operatorname*{e}\nolimits^{\left(  \Gamma_{c}%
+i\chi\right)  t/2}a$. Next, we proceed to the third approximation $iii)$
noting that for the time interval of the atom-field interaction, about
$10^{-4}$s, and for the spontaneous-emission decay times of circular Rydberg
states $\Gamma_{a}\sim10^{2}$ s$^{-1}$\cite{Haroche01}, the second terms on
the right hand side of Eqs. (\ref{E4a}) and (\ref{E4b}) can be dismissed.
Therefore, within the above approximations we finally obtain the solution%
\end{subequations}
\begin{equation}
a=\operatorname*{e}\nolimits^{-(\Gamma_{c}+i\chi)t/2}\left(  a_{0}%
\cosh\widetilde{r}_{on}+\operatorname*{e}\nolimits^{-i\varphi_{on}}%
a_{0}^{\dagger}\sinh\widetilde{r}_{on}\right)  \text{,} \label{E5}%
\end{equation}
from which we obtain the variance in the squeezed quadrature ($\varphi
_{on}=\pi/2$)
\begin{equation}
\left\langle \Delta X\right\rangle ^{2}=\frac{1}{4}\left[  1-\left(
1-\operatorname*{e}\nolimits^{-2\widetilde{r}_{on}}\right)  \operatorname*{e}%
\nolimits^{-\Gamma_{c}t}\right]  \label{E6}%
\end{equation}
where the squeezing factor under the atomic decay is $\widetilde{r}%
_{on}=4\left|  \xi\right|  (1-\operatorname*{e}\nolimits^{-\Gamma_{a}%
t/2})/\Gamma_{a}$. As noted after Eqs. (\ref{E4a}) and (\ref{E4b}), for the
time interval of the atom-field interaction $\operatorname*{e}%
\nolimits^{-\Gamma_{a}t/2}\approx1-\Gamma_{a}t/2$, such that the squeezing
factor under atomic decay $\widetilde{r}_{on}$ is approximately that of the
ideal case $r_{on}$. However, the damping of the cavity mode, expressed by the
time-dependent exponential decay in Eq. (\ref{E6}), contributes substantially
to increase the variance of the squeezed quadrature and, consequently, to
decrease the squeezing rate. Assuming the decay time of circular Rydberg
states $\Gamma_{a}\sim10^{2}$ s$^{-1}$ (when $n\approx50$) and the typical
values considered above for the parameters $\left|  \lambda_{g}\right|  $,
$\left|  \lambda_{e}\right|  $, $\left|  \Omega\right|  $, $\delta$, and $t$,
we obtain $\widetilde{r}_{on}\sim1.06$. Therefore, for the typical decay
factor for open high-$Q$ cavities, $\Gamma_{c}\sim10^{3}$ s$^{-1}$
\cite{Haroche01}, we obtain the variance in the squeezed quadrature
$\left\langle \Delta X\right\rangle ^{2}\sim7\times10^{-2}$, representing a
squeezing around $72\%$. For closed high-$Q$ cavities, where $\Gamma_{c}%
\sim10$ s$^{-1}$ and noncircular Rydberg levels with $n\sim60$ are employed,
such that, $\Gamma_{a}\sim5\times10^{3}$ s$^{-1}$ \cite{Walther}, we obtain
$\left\langle \Delta X\right\rangle ^{2}\sim4.7\times10^{-2}$ and a squeezing
around $81\%$. We note that in closed cavities an external amplification field
directly coupled to a second normal mode of the cavity could be used
\cite{Solano}.

There are others sensitive points in the experimental implementation of the
present scheme. Apart from the atomic detection efficiency and the spread of
the atomic velocity not taken into account in the present analysis, the
Gaussian profile $f(x)$ of the cavity field in the transverse direction must
also be computed. Due to this Gaussian profile the atom-field couplings
$\lambda_{g}$ and $\lambda_{e}$ becomes time-dependent parameters as well as
the effective amplitude of the parametric amplification field which turns to
be (without considering dissipation) $\xi=2\Omega\lambda_{a}\lambda_{b}\left[
f(x)\right]  ^{2}/\delta^{2}$ where $f(x)=\exp(-x^{2}/w^{2})$ ($x$ being the
time-dependent atom position from the center of the cavity, and $w\sim0.6$ cm
\cite{Haroche01} is the waist of the Gaussian). The effect of the field
profile can be evaluated by using the analytical results for a time-dependent
parametric amplification process, demonstrated in \cite{Celso}, leading to the
on-resonant squeezing factor $r_{on}^{\prime}=2\int_{0}^{\tau}\xi(t)dt=\left(
\left.  4\Omega\lambda_{g}\lambda_{e}\right/  \delta^{2}\right)  \int
_{0}^{\tau}\left[  f(x)\right]  ^{2}dt$. Considering the atom-field
interaction time about $\tau\sim2\times10^{-4}$s, we get the squeezing factor
$r_{on}^{\prime}\sim0.4$ representing $\left\langle \Delta X_{1}\right\rangle
^{2}\sim1.1\times10^{-1}$ and a squeezing around $55\%$. To obtain the value
$r_{on}^{\prime}\sim1$ of the ideal case, we must increase the interaction
time to $\tau\sim5\times10^{-4}$s. However, with this value of the atom-field
interaction time the dissipative process becomes more pronounced and a
coast-benefit estimative must be computed. A detailed analysis involving both
error sources, the dissipative process and the Gaussian profile of the cavity
field will be considered elsewhere \cite{Villas}.

In conclusion, we have shown theoretically that the dispersive interaction of
a cavity mode prepared in the state $\left|  \Psi\right\rangle $ with a driven
atom would produce the squeezing operation $S(\zeta)\left|  \Psi\right\rangle
$. In the ideal case we would obtain squeezing around 88\% of a prepared
coherent field state, in the on-resonant regime, with the passage of a single
three-level atom through the cavity. We finally stress that the squeezing of
previously prepared states is crucial to build truly mesoscopic superpositions
with a large average photon number produced by the parametric amplification
process we have engineered \cite{Celso}.

\begin{acknowledgments}
We wish to express thanks for the support from FAPESP (under contracts
\#99/11617-0, \#00/15084-5, and \#02/02633-6) and CNPq (Instituto do
Mil\^{e}nio de Informa\c{c}\~{a}o Qu\^{a}ntica), Brazilian agencies.
\end{acknowledgments}

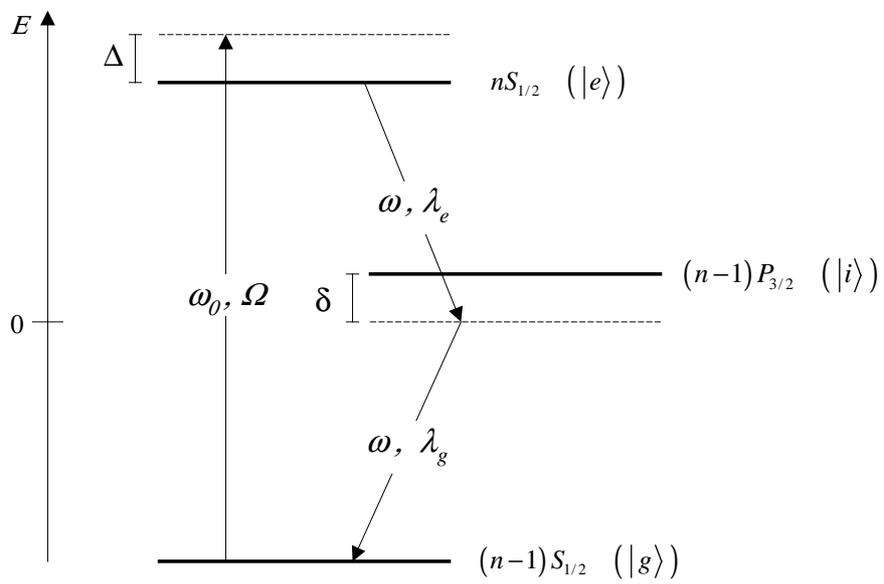
Fig. 1. Energy-levels diagram of the three-level atom for the parametric
amplification scheme.

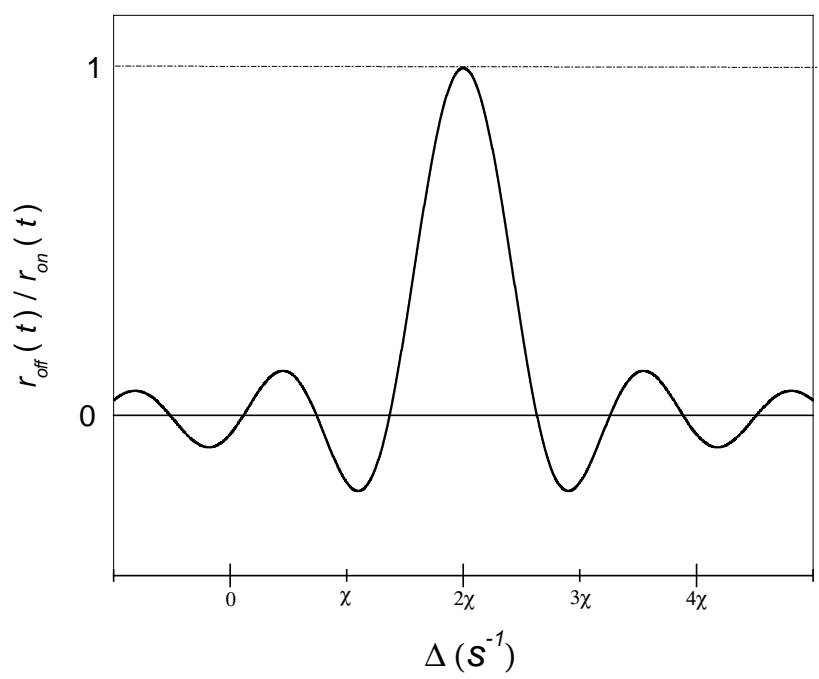
Fig. 2. Ratio of the squeezing factors in the off-resonant and on-resonant regimes.
\end{document}